\newcommand{\be}{\begin{eqnarray}}
\newcommand{\ee}{\end{eqnarray}}
\newcommand{\nn}{\\ \nonumber}

\newcommand{\vecg}[1]{\mbox{\boldmath $#1$}}

\documentclass[12pt]{article}

\usepackage{amsmath,amssymb}
\usepackage{graphicx}
\usepackage{cite}
\usepackage{bm}
\begin{document}

\vskip 2cm

\begin{center}

{\Large Super-Yang-Mills quantum  mechanics  and      supermembrane
spectrum} 
\footnote{Bern preprint BUTP-89/24 published in the Proceedings 
 of the conference on Supermembranes and Physics in 2+1 Dimensions 
(Trieste, 1989).}

\vskip 4cm

A.V. \textsc{Smilga} \\
\texttt{smilga@subatech.in2p3.fr}

\end{center}

\vskip 5cm

\begin{abstract}

It is shown that the mass spetrum of supermembrane theory is continuous. 
This fact is due to the trivial circumstance that membrane can emit "needles" 
of zero area with no cost in energy. 
In supersymmetric case, this classical degeneracy 
is not lifted by  quantum corrections. This unpleasant property may be cured, 
perhaps, 
for the supermembrane with a modified action involving higher derivative terms.

\end{abstract}

\newpage

\section{Introduction}

    Recently the impressive progress in studying string theories which 
impelled many high energy theorists to believe that the Theory of Everything 
based on string philosophy would be built soon has slowed down. The task 
proved to be more complicated than it seemed to be several years ago and a 
decisive breakthrough has not occurred yet. In these circumstances, the 
attention of theorists was distracted to studying  alternative 
possibilities to construct the theory of fundamental interactions. 
One of these alternative candidates is the supermembrane theory - 
the main subject of the present meeting.
   The main conclusion of my talk is that, unfortunately, the  standard  
supermembrane  with  the  action  being  a
supersymmetry extension of the Nambu action in 2+1 dimensions
meets serious difficulties. 
It involves the continuous mass spectrum and hence, unlike string theory, 
the effective field theory of lowest mass states cannot be built in 
supermembrane case.
   This unpleasant fact has a very transparent physical interpretation . 
In the light cone gauge, the supermembrane mass operator is an integral 
of some density over membrane surface (see Refs.\cite{1,2} and the
lectures by   I.Hoppe, C.Pope
and others at this Conference) :
   \be
M^2_{\rm supermembrane} \ =\ \int d^2\sigma \left[
P'^2_I + \frac 12 \{X_I, X_J\}^2 + 2i\bar \theta \Gamma_I 
\{\theta, X_I\} \right] \, ,
  \ee
where $I,J = 1,\ldots, 9; \ \{X_I, X_J\} = 
\epsilon^{rs} \partial_r X_I \partial_s X_J$ and $P'_I$ 
involves only nonzero modes contribution. 
But the membrane (unlike string) can be deformed without increasing its area: 
a "needle" of zero area can be attached at any point of membrane surface 
(see Fig.1).

\begin{figure}[h]
   \begin{center}
 \includegraphics[width=2.5in]{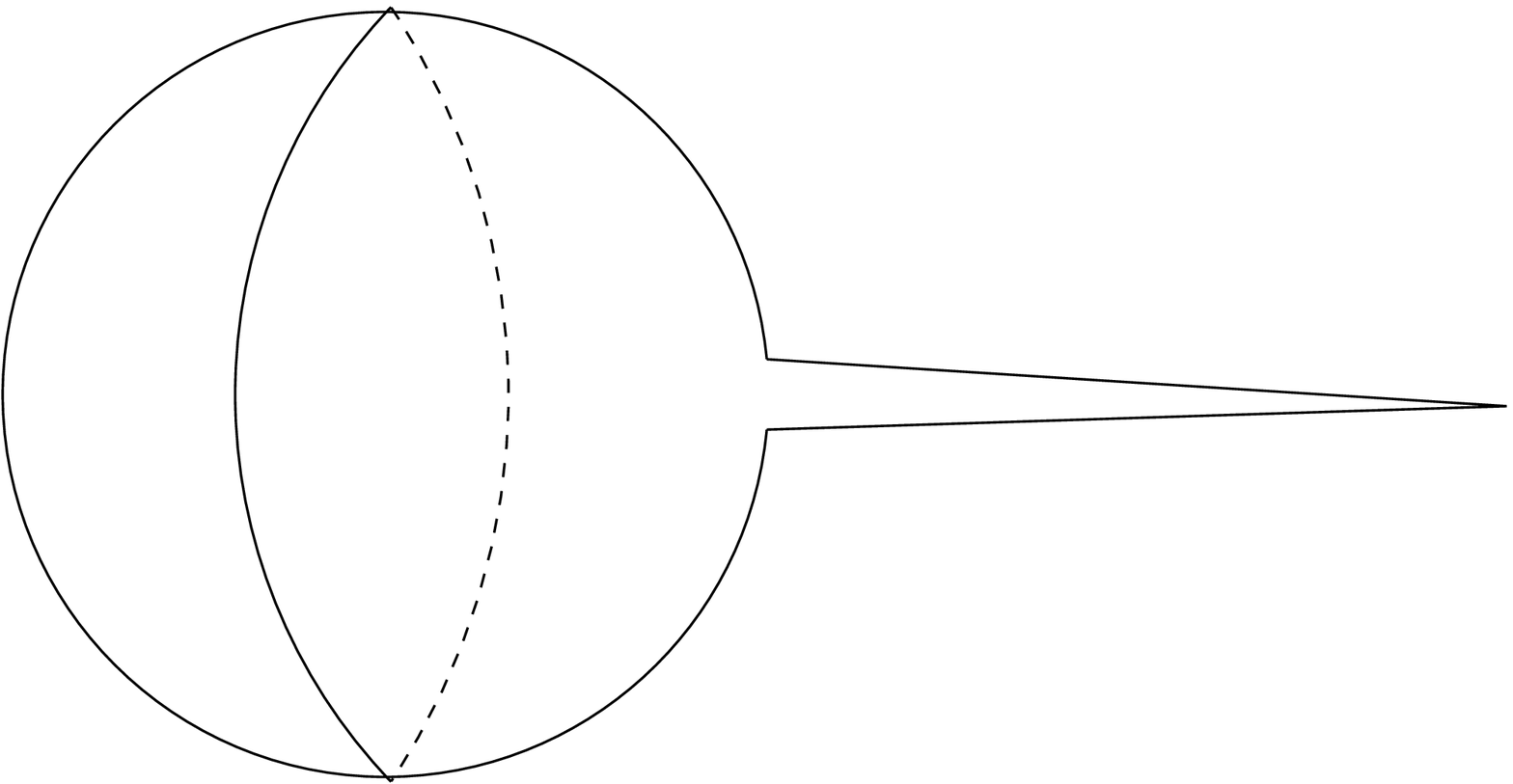}
\caption{}
    \end{center}
\end{figure}

The classical energy of these degenerate configurations is the same. 
Thus, we are faced with the valley in configuration space, the motion 
along which is infinite, and the spectrum is continuous.
The statement is that, in supersymmetric case, this
classical degeneracy is not lifted by quantum corrections. 
We use the wonderful fact discovered in  Ref.\cite{2} that the
supermembrane hamiltonian (1) may be thought of as $N \to \infty$
1imit of the hamiltonian of the Super-Yang-Mills quantum mechanics 
with the $SU(N)$ group,
   \be
  2H^{\rm SYM\ QM} \ =\ E^A_I E^A_I + \frac 12 f_{ABE} f_{CDE} 
A^A_I A^B_J A^C_I A^D_J + 2i f_{ABC} \bar \lambda^A \Gamma_I \lambda^B A^C_I 
 \, ,
  \ee
where $f_{ABC}$   are the $SU(N)$ structure constants, $A^A_I$ and 
$\lambda_\alpha$   are the
 boson and fermion components of supergauge field, $\Gamma_I$
are the
10-dimensional $\Gamma$-matrices. The hamiltonian  (2)    is obtained by 
reducing the 10-dimensional SYM theory to (0+1) space (i.e. by considering only the
 fields not depending on the space variables) and choosing the gauge 
$A^A_0= 0$. The correspondence with the supermembrane hamiltonian is the 
following: $A^A_I$ correspond to the spherical harmonics of the coordinates
   \be
  X_I(\sigma) \ =\ \sum_{LM} (X^{LM}_I \equiv A^A_I ) Y^{LM} (\theta,\phi) \, ,
   \ee
$P'^{LM}_I \equiv E^A_I$, $\theta_\alpha^{LM} \equiv \lambda_\alpha^A$.
There is a subtlety that, at finite $N$, the
structure constants of $SU(N)$ do {\it not} coincide with the structure 
constants of the area-preserving diffeomorphism group which, in fact, enter 
eq. (1), but only tend to them in the $N \to \infty$  limit.
 But for the physical implications, the property
 \be
\lim_{N \to \infty} f_{ABC}^{SU(N)} \ =\ g_{ABC}^{{\rm SDiff}^2} 
   \ee
is quite sufficient. One may think of the hamiltonian (2) as  of  a  
convenient  way  to  perform   ultraviolet regularization of the 2-dimensional 
field theory (1). It is very close in spirit to familiar lattice 
regularization of YM theory where one cares to respect the gauge symmetry 
of the field theory. As $N  \to   \infty$         and we involve in the 
analysis the spherical harmonies with higher and higher L's, the 
physical properties of the hamiltonians of eq. (2) and of eq. (1) are 
equivalent.
    The hamiltonian (2) involves valleys, i.e. the regions in configuration 
space where the classical bosonic potential turns to zero. A sufficient 
(but for $N > 2$ not necessary - see the discussion in Sect.3) condition of 
the valley is
  \be
  A_I^{A({\rm val})} = c_I \eta^A \, . 
  \ee
In that case, the commutator  $[A_I, A_J]$ together with the
potential  $\propto {\rm Sp} \left\{[A_I, A_J]^2 \right\}$ 
turn to zero. The existence of
these valleys and also the fact that, in supersymmetric case, they are {\it not} 
lifted by quantum corrections was first noted
by E.Witten in the celebrated paper 
where the concept of the Witten index had been introduced [3].

   Witten was interested in supersymmetric qauqe field theories and studied 
the question if the supersymmetry is broken spontaneously. To this end, he 
considered a theory
placed in a small spatial box with $L \ll \Lambda$ so that $g^2(L) \ll  1$ 
and perturbative expansion makes sense, and imposed periodic boundary 
conditions on the fields
  \be
\Phi (x_1, x_2, x_3) = \  \Phi(x_1 +L , x_2, x_3 )                 
  \ee                  
 etc.   (Witten considered also t'Hooft twisted boundary conditions but, for our
 purposes, they are not relevant). In this case, only the gauge transformations 
respecting the boundary conditions (6) are allowed and, in contrast to the field
 theory in infinite volume, the field configuration
$A^A_I$        = const    is  generally  {\it not}      gauge  equivalent  to
the
configuration  $A^A_I = 0$. Thus, for the gauge theory in finite
volume, the valley of constant configurations $A^A_I$ satisfying
the condition $[A_I, A_J] =0$   is present. As long as $L$ is
nonzero, the valley has {\it finite} length $\propto   2n/gL$ (the
configurations $A^3_I (\vecg{x})$ and $A^3_I (\vecg{x})  + 2\pi n /gL$,
 where $n_I$
  is an
integer vector, are gauge equivalent \cite{3} ).  But at small $L$,
 the valley length is large and in the supersymmetric case, where, 
as Witten argued,
 the valleys are preserved after accounting for the quantum corrections,
 the low 
energy spectrum of the system is determined by  the motion along the valleys: 
the
characteristic excitation energies due to the valley motion
are $ g^2(L)/L$,  which is much less than the characteristic
energies of "fast variables" excitations  $\propto 1/L$.
   To resolve the problem at hand, one should go one step further and set 
$L = 0$. In this case (where field theory is reduced to quantum mechanics),
 the valleys have infinite
length and the spectrum of the corresponding hamiltonian is continuous.
    In the rest of the talk, I'11 try to argue why, in the supersymmetric case,
 the valleys are not lifted. I start with the analysis of the simple case of 
quantum mechanics
of $d=4$ $SU(2)$ SYM theory, which is presented in the following section. 
In Sect.3, I consider its extension to higher dimensions and higher groups. 
In Sect.4, I'll discuss briefly chiral gauge theories where the effective 
valley hamiltonian is not so trivial. Conclusive remarks are given in the last
 section.

\section{
Quantum mechanics of $SU(2)\  d =  4$ super-Yang-Mills}

   The hamiltonian and supercharges of the model have the form \cite{4,5} 
  \be
H \ =\ \frac 12 P^A_i  P^A_i + \frac {g^4}4 \epsilon^{ABE}  \epsilon^{CDE} 
A^A_i A^B_j A^C_i A^D_j + ig \, \epsilon^{ABC} \bar \lambda^A \sigma_i 
\lambda^B A^C_i \, ,
\ee
 \be
Q_\alpha  \ =\ \frac 1{\sqrt{2}} (\sigma_j)_\alpha^{\ \gamma} \lambda_\gamma^A
(E^A_j + i H^A_j) \, ,  \nn
\bar Q^\beta  \ =\ \frac 1{\sqrt{2}} (\sigma_j)_{\ \delta}^{\beta} \bar 
\lambda^{\delta B}
(E^B_j - i H^B_j) \, ,
  \ee
where $i,j  = A , B  =  1,2 ,3 ; \ \alpha =  1,2$  and $H^A_I  = \frac g2 
\epsilon^{ABC} \epsilon_{ijk} A^B_j A^C_k$. 
We introduced the coupling constant $g$ which can be easily
scaled away by the transformation $A  = g^{-1/3} \tilde{A}, 
\ \lambda = g^{-1/3} \tilde{\lambda}$.
The supercharges (8) satisfy the following supersymmetry algebra
   \be
\{Q_\alpha, \bar Q^\beta \} \ =\ \delta_\alpha^\beta H - 
(\sigma_i)_\alpha^\beta A^A_i G^A \, ,
\ee
where
   \be
G^A \ =\ \epsilon^{ABC} (P^B_j A^C_j + 
i \bar \lambda_\alpha^B \lambda_\alpha^C)
  \ee
is the Gauss constraint operator. The second term in the
r.h.s. of eq. (9) is due to the known fact that a superposition of two 
supertransformations involves a gauge transformation besides translation. 
In the Hilbert space
of physical gauge-invariant states,  $G^A |\Psi_{\rm phys} \rangle = 0$ 
 and the
standard $N = 2$ SQM algebra is realized.
    The hamiltonian (7) involves the valleys defined by the condition (5) 
(in the $SU(2)$ case, it is both sufficient and
necessary). The $A^{\rm A (val)}_i$
of Eq. (5) can be freely rotated in
 colour space. This is an unphysical gauge degree of freedom,
 and one is allowed to choose the gauge where  $\eta^A = \delta^A_3$.     
  In  the Born-Oppenheimer spirit, 
 we may classify the physical bosonic variables
in two groups: the "slow"
variables $c_i$ which describe the motion along the valley and
the
"fast" variables $A^a_i$, where $a =  1,2$ and  $A^a_i c_i  =  0$. 
Only three of                                                               
four variables $A^a_i$  are physical, while the fourth is the
gauqe degree of freedom corresponding to a gauge rotation around the third 
colour axis. Our task is to construct the effective hamiltonian depending 
only on the slow variables
$c_i$ and describing the low-energy spectrum of the system. To
this end, we assume $ |\vecg{c}|   \gg     |\vecg{A}^a|$  and classify the 
various
terms in the full hamiltonian (7) by the powers of the formal parameter 
$x^{\rm fast}/x^{\rm slow} $.  We get
  \be
 H \ =\ H^{(0)} +  H^{(2)} +  H^{(2)} \, ,
  \ee 
where
   \be
   H^{(0)} \ =\ - \frac 12 \frac {\partial^2 }{(\partial A^a_m)^2}
+ \frac {g^2 c^2}2 ( A^a_m A^a_m) + igc \, \epsilon^{ab} \bar\lambda^{a \alpha}
(\sigma_3)_\alpha^\beta \lambda^b_\beta
 \ee
(we have chosen the $z$ direction along $\vecg{c}$  so that 
$c_i = c\delta_{i3}$ 
and $a,m =   1,2$ ) and $H^{(1)}  , H^{(2)}$   involve cubic and quartic in
$A^a_m$ terms, correspondingly.

   The main observation is that the hamiltonian (12) describes the 
supersymmetric oscillator. Its ground state can be found explicitly:
   \be
\Psi_C^{(0)} (A^a_m) \ \propto \ 
\exp\left\{ - \frac {gc}2 A^a_m A^a_m \right\}
\left\{ \lambda^{b\alpha} \lambda_\alpha^b + i\epsilon^{bc} \lambda^{b\alpha}
(\sigma_3)_\alpha^\beta \lambda^c_\beta \right\}\, .
 \ee
It satisfies automatically the constraint   $G^3 |\Psi^{(0)}_C \rangle  =  0$
and has zero energy. In other words, in supersymmetric case,
 zero-point quantum 
fluctuations of the fast variables in the direction across the valley yield 
zero contribution in the effective hamiltonian.
   Note that, in the nonsupersymmetric case, it is not true. For the pure YM 
quantum mechanics, $ H^{(0)}$ does not involve the third term and the 
bosonic zero-point energy  $\sim gc$ has no fermionic counterpart to cancel
 with. Thus, in the pure
bosonic case, the classical valleys are lifted by quantum corrections, 
the motion along the valley is prohibited and the spectrum is discrete. 
A clear illustration what is
going on in the pure bosonic case is given in Fig.2. 
As $c$ grows, the walls of the valley get steeper and steeper and the 
zero-point energy of transverse fluctuations gets higher and higher.

\begin{figure}[h]
   \begin{center}
 \includegraphics[width=2.5in]{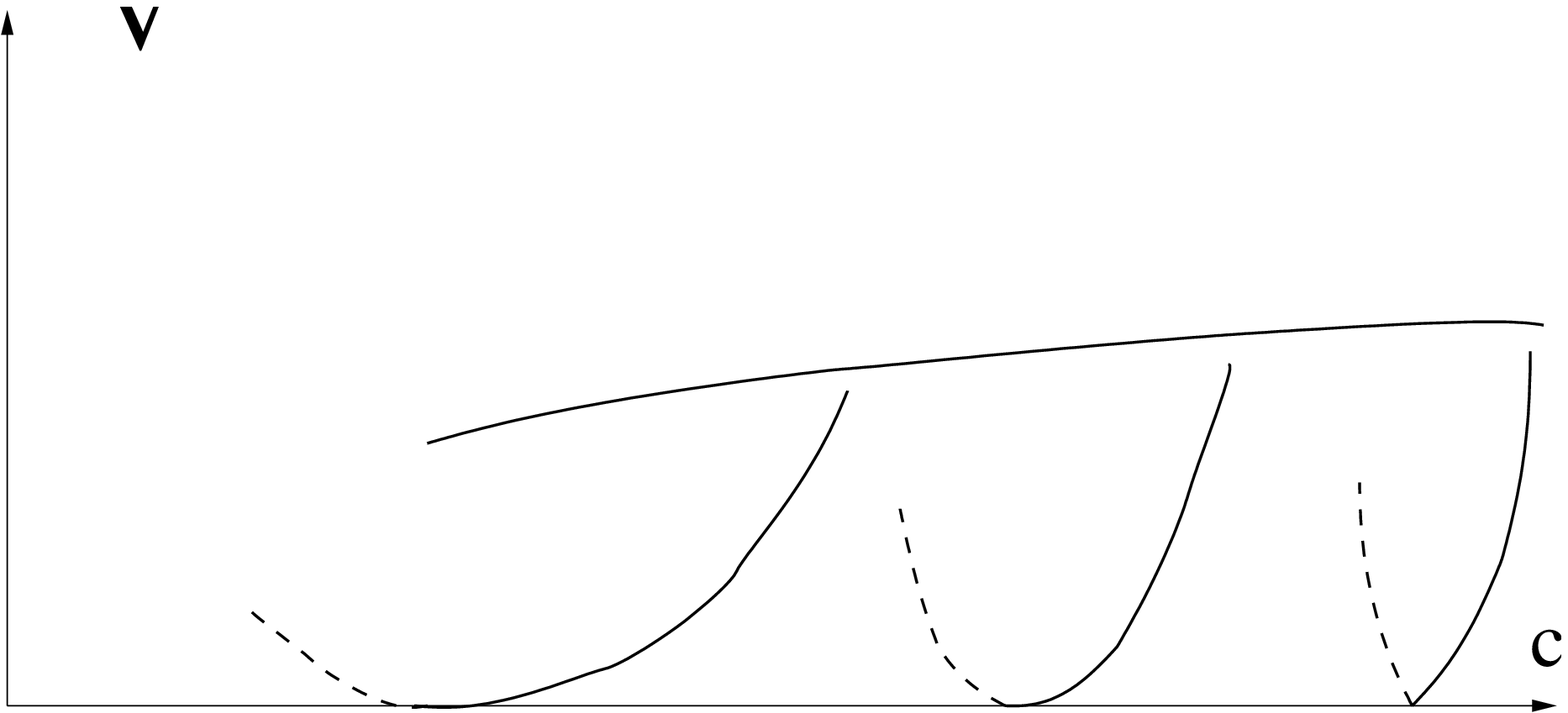}
    \end{center}
\caption{}
\end{figure}

   Surely, the effective valley hamiltonian can be obtained in the rigorous 
and regular way. The straightforward way to do it is to use second order 
perturbation theory in the
parameter $x^{\rm fast}/x^{\rm slow} $ 
and to write
  \be
H^{\rm eff} \ =\ \langle 0 | H^{(2)} | 0 \rangle - 
\sum_n \frac { \langle 0 | H^{(1)} | n\rangle  \langle n | H^{(1)} | 0 \
\rangle}{E_n} \, ,
  \ee
where $|0\rangle$  and $|n \rangle$ 
 are the ground and excited states of the hamiltonian $H^{(0)}$.
 More  simple  and more  convenient  is to find out first the {\it effective 
supercharge}
  \be
Q_\alpha^{\rm eff} \ =\  \langle 0 | Q_\alpha | 0 \rangle
\ee
 with  $Q_\alpha$  of  eq. (8),  and  then  to  built  up
$H^{\rm eff}$  as
the anticommutator of $Q_\alpha^{\rm eff}$
and $\bar{Q}^{\alpha \, {\rm eff}} $.
 Both methods give
the same answer
   \be
Q_\alpha^{\rm eff} \ =\ -i \frac 
{(\sigma_k)^\beta_\alpha \lambda_\beta}{\sqrt{2}} \, \frac {\partial }
{\partial c_k}, \nn
H^{\rm eff} \ =\ - \frac 12 \frac {\partial^2}{(\partial c_i)^2} \, ,
  \ee
i.e. the motion along the valley is unbounded, indeed.

   Let us discuss now the region of applicability of these results. The 
characteristic values of $A^a_i$ in the wave
function  (13) are    $(A^a_i)_{\rm char} \sim      1/\sqrt{gc}$.
    The condition  $(A^a_i)_{\rm char} \ll  c$
 necessary for the classification (11)   to make
sense is fulfilled provided $gc^3   \gg 1$. The parameter $1/gc^3$   is
the true Born-Oppenheimer expansion parameter in the effective hamiltonian. 
If $gc^3 \gg 1$,  the corrections to the lowest order effective hamiltonian 
(14) are expected to be small.
   We calculated explicitly these corrections in the case of supersymmetric 
QED, which is technically more simple [6].
Supersymmetric QED involves the Abelian gauge field $A_i$, its
supersymmetric counterpart $\lambda_\alpha$, and two chiral multiplets
$(\phi_\alpha, \xi_\alpha)$    and $(\chi_\alpha, \eta_\alpha)$
   with opposite charges. The slow
valley variables in this case are just $A_i$
and the fast variables are $\phi$ and $\chi$. 
 The result of calculation of the effective hamiltonian for the massless SQED 
quantum mechanics is
  \be
H^{\rm eff} \ =\ -\frac 12 f \stackrel {\longrightarrow}
{(\partial^2/\partial A_k^2) f} + i\epsilon_{kpl} 
\bar \lambda \sigma_l \lambda\, f (\partial f/\partial A_p) \frac{\partial}{\partial A_k}
+ \frac 16 f (\partial^2 f/\partial A_k^2) \, \bar \lambda \sigma_l \lambda 
\, \bar \lambda \sigma_l \lambda \, ,
  \ee
where
  \be
f(\vecg{A}) \ =\ 1 - \frac 1{4eA^3} 
 \ee
(the corresponding effective supercharges may be found in Ref. \cite{6}).
We see that, as long as $eA^3 \gg  1$, the second term in the
r.h.s. of
eq.(18)
is small and $H^{\rm eff}$ 
is reduced
to the free
hamiltonian in eq.(16). But in the region where A is small, the corrections 
grow large and the Born-Oppenheimer approximation is not applicable.
   However, for the conclusion that the spectrum of $H^{\rm eff}$ and hence the 
low energy spectrum of the total hamiltonian is continuous this is not 
important as this conclusion depends only on the fact that, at large values
 of A, the motion is free.

\section{Higher groups and higher dimensions}

   For supermembrane,
 relevant is the quantum mechanics of 10-dimensional $SU(N)$ Super-Yang-Mills.

   Consider first the SYM theory with the $SU (2)$ gauge group in 10 dimensions. 
One can be convinced that this case can be treated quite analogously to the 4-dimensional case. 
We have now 9 valley variables $A^3_I = c_I$  and 16 fast variables 
$A^a_M$  ( $a=  1,2$ and $M =  1,\ldots,8$), one of which corresponds to the
gauge degree of freedom connected
with rotations around the
third colour axis. Directing $c_I$
along
the  9-th axis: $c_I   = c\delta_{I9}$ 
 and  assuming  $gc^3      \gg 1$,  we  can  expand  the
total
hamiltonian
over the
formal parameter  $x^{\rm fast}/x^{\rm slow} $ . 
Then
the analog of $H^{(0)}$ is 
  \be
   H^{(0)}_{\rm 10\ dim} \ =\ - 
\frac 12 \frac {\partial^2 }{(\partial A^a_M)^2}
+ \frac {g^2 c^2}2 (A^a_M A^a_M) + igc \,\epsilon^{ab} \bar\lambda^{a \alpha}
(\Gamma_9)_\alpha^\beta \lambda^b_\beta \, ,
 \ee
$\alpha, \beta = 1, \ldots,8$.  It is still a supersymmetric oscillator
 with the ground state wave function
\be 
\Psi_C^{(0)} (A^a_M) \ \propto \ 
\exp\left\{ - \frac {gc}2 A^a_M A^a_M \right\}
\left\{ \lambda^{b\alpha} \lambda_\alpha^b + i\epsilon^{bc} \lambda^{b\alpha}
(\Gamma_9)_\alpha^\beta \lambda^c_\beta \right\}^4\, ,
 \ee
which has zero energy. All the arguments of the previous
section are repeated unchanged with the result that the lowest order effective
 hamiltonian describes the free motion in
the $c_I$ - space
and, in the region $gc^3 \gg  1$, the corrections are small.

Let us discuss now  a slightly more complicated case
of higher $SU(N)$    groups.
Consider first the $SU(3)$    case. The
valley condition $[\hat A_I, \hat A_J] =0 $
   is satisfied provided  all $\hat A_I$ can be simultaneously diagonalized
by a gauge transformation:
  \be
\hat A_I^{\rm (val)} \ =\ \left( 
\begin{array}{ccc} 
a_I & 0 & 0 \\ 0&  b_I - a_I & 0 \\ 0 & & -b_I 
\end{array}
\right) \, .
  \ee
This representation is
more convenient than the equivalent
form $ \hat A_I =     A^3_I t^3 + A^8_I t^8$. 
Thus, in the $SU (3 )$    case, the valley         
variables are $a_I$ and $b_I$                
- the weights of the Cartan subalgebra.
The valley configuration (21) is not presented, generally,
in the form (5) - the vectors $a_I$ and $b_I$ are not necessarily    parallel. 
As earlier, we substitute now $\hat A_I = \hat A_I^{\rm val} + \hat A_I^{\rm 
fast}$ 
  in the hamiltonian (2) and pick up the quadratic
in $\hat A_I^{\rm 
fast}$ terms. Then the potential part of $H^{ (0)}$ 
is
   \be
V^{(0)} \ =\ \frac {g^2(2\vecg{a} - \vecg{b})^2}2 A^{a=1,2}_I A^{a=1,2}_J
\left[ \delta_{IJ} - \frac{(2a-b)_I (2a-b)_J }{(2\vecg{a} - \vecg{b})^2} \right]
+  \nn 
\frac {g^2(\vecg{a} + \vecg{b})^2}2 A^{a=4,5}_I A^{a=4,5}_J
\left[ \delta_{IJ} - \frac{(a+b)_I (a+b)_J }{(\vecg{a} + \vecg{b})^2} \right]
+ \nn
 \frac {g^2(2\vecg{b} - \vecg{a})^2}2 A^{a=6,7}_I A^{a=6,7}_J
\left[ \delta_{IJ} - \frac{(2b-a)_I (2b-a)_J }{(2\vecg{b} - \vecg{a})^2} \right]
\, .
   \ee

   Thus, there are 48 fast variables divided naturally in three groups:
$ A_I^{1,2}$   satisfying the condition $(2\vecg{a}- \vecg{b}) \vecg{A}^{1,2} 
= O$,   $ A_I^{4,5}$   satisfying the condition   
$(\vecg{a}+ \vecg{b}) \vecg{A}^{4,5} = O$   and $ A_I^{6,7}$
satisfying the condition $(2\vecg{b} - \vecg{a}) \vecg{A}^{6,7} = O$.
Two of these 48
variables are gauge degrees of freedom corresponding to the
action of the generators $G^3$   and $G^8$.
Besides, there are 6 more
gauge
degrees  of  freedom : $ (A_I^{1,2})_{\rm gauge} = 
X^{1,2}_{\rm gauge} (2a-b)_I  $;
 $ (A_I^{4,5})_{\rm gauge} = X^{4,5}_{\rm gauge} (a+b)_I  $ and 
 $ (A_I^{6,7})_{\rm gauge} = X^{6,7}_{\rm gauge} (2b-a)_I  $
corresponding to the rotations generated by $G^{1,2,4,5,6,7}$ that
act nontrivially on the valley configuration (21).

   Each of the three terms in the r.h.s. of eq. (22) combined with the 
corresponding terms in the kinetic part of $H^{(0)}$ and the part involving 
fermions has the form of eq.(19). 
Thus, in the $SU(3)$ case, $H^{(0)}$  is represented as the sum of three 
hamiltonians, each of them having the same form as in the $SU ( 2 )$
 case, i.e. representing supersymmetric oscillator with zero
enerqy ground state of eq .(20). The total ground state wave
function representing the product of three eq.(20)-like
factors satisfies automatically  the  conditions $G^3| \Psi_0 \rangle = 
G^8| \Psi_0 \rangle = 0$. 
 The effective supercharges and hamiltonian can
be found in the same way as earlier. They have the form
 \be
Q_\alpha^{\rm eff} \ =\ - \frac i{\sqrt{2}} 
(\Gamma_I)^\beta_\alpha \left[ \lambda_\beta^{(a)} \frac {\partial}
{\partial a_I} + \lambda_\beta^{(b)} \frac {\partial }
{\partial b_I} \right] \, , \nn
H^{\rm eff} \ =\ - \frac 23 \left[ \frac {\partial^2}{(\partial a_I)^2}
+  \frac {\partial^2}{(\partial b_I)^2} + 
\frac {\partial^2}{\partial a_I \partial b_I} \right]\, .
  \ee
The hamiltonian in (23) describes the free motion in the space
of weights $a_I , b_I$ 
(there is a subtlety that the wave
function  is  required  to  be  invariant  under  Weyl
transformations permuting weights but, for our purposes, it is not important). 
The corrections to the lowest order $H^{\rm eff}$
 of eq. (23) are small provided $ g | 2\vecg{a}- \vecg{b} |^3, g|\vecg{a} + 
\vecg{b} |^3, g| 2\vecg{b}- \vecg{a}|^3  \gg 1$
, and the spectrum is continuous.

   The higher $N$ case is treated with an equal ease. Thus, for $SU(4)$,
 we have 3x9  = 27 valley variables $a_I , b_I , c_I$   and
96 fast variables divided in 6 groups, which include three
gauge degrees of freedom corresponding to $G^3 , G^8$
and $G^{15}$
rotations; for $SU (5)$ there are 4x9 = 36 valley variables and 
160 fast variables divided in 10 groups, etc. 
The answer is the same: $H^{\rm eff}$ describes the free motion in the space of 
weights and the spectrum is continuous.
    The similar analysis with the same result has been done in Ref. \cite{7}, but in 
the particular case where the valley configuration is representable in the form 
(5), i.e. when all the weights vectors $\vecg{a}, \vecg{b}, ...$ are parallel.
   In this respect, their treatment is not quite complete,  though 
it is quite
sufficient to justify the continuity of the spectrum.

\section{Digression: chiral gauge theories}

In some cases, the lowest order effective hamiltonians are not so trivial as 
those in eqs.(16,23) and are rather funny. Nontrivial contributions 
arise for gauge theories with chiral matter content - the only case not 
analyzed in  Witten's paper [3]. In our works \cite{8,9},
 the effective hamiltonians for chiral supersymmetric electrodynamics 
(to be anomaly free, the charges of different chiral multiplets
should satisfy the condition $\sum_f (Q_f)^3 = 0$), 
the chiral supersymmetric $SU (3)$ theory involving 
a right sextet and 7 left triplets, and the chiral $SU (5)$
 theory with a left quintet and a right decuplet has been built. 
We have found effective hamiltonians
 both for field theories in a finite volume 
(in the spirit of Witten's approach)
 and in the zero-volume (quantum mechanical) limit.
   We present here the supercharges and hamiltonian for quantum mechanics of 
chiral SQED with one chiral multiplet
found in Ref. \cite{8} 
(in the quantum mechanical limit,  the problem of anomaly does not arise). 
They  depend  on  the  slow
variables $A_i , \lambda_\alpha$   and have the form
   \be
Q_\alpha^{\rm eff} \ =\ - \frac i{\sqrt{2}} \lambda_\gamma
\left[ 
(\sigma_j)^{\ \gamma}_\alpha \left( \frac {\partial}
{\partial A_j} - i{\cal A}_j \right) + \frac {\delta_\alpha^\gamma}{2A}
\right], \nn
H^{\rm eff} \ =\ - \frac 12 \left( \frac{\partial}
{\partial A_j} - i{\cal A}_j \right)^2 + \frac 1{8A^2} + \frac {A_j}{8A^3} 
\bar \lambda \sigma_j \lambda \, ,
  \ee
where $\vecg{\cal A}$   is a vector function of the fields $\vecg{A}$
coinciding
with the Dirac monopole vector potential. The hamiltonian in (24)
describes the motion of scalar (in the sectors F = 0,2) and spinor 
(in the sector F = 1) particle in the field of a magnetic monopole
 placed at the point $\vecg{A} = 0$ and also in the
scalar potential $ 1/8A^2$.  The spectrum of the hamiltonian in (24) is still 
continuous as, at large A, the potentials
(both the scalar and the vector) and the monopole magnetic field 
${\cal H}_j = 
-A_j /2A^3$   vanish. (The continuity of spectrum depends only on
two premises: {\it (i)} the existence of unbounded valleys and 
{(ii)} supersymmetry implying the zero ground state energy of the 
fast variables hamiltonian $H^0$. )   The Born-Oppenheimer
parameter is $1/eA^3$     as earlier and, at small   $A$,   the
corrections are large and the Born-Oppenheimer approximation is not applicable.
    
There  is a gauge  freedom  in the  phase choice  of
eigenfunctions of the hamiltonian in (24). This phenomenon is closely 
connected to the singularity of $H^{\rm eff}$ at $\vecg{A} = 0$
 where the energy gap due to the fast variables excitations vanishes
and is known in the literature as Berry's phase \cite{10}.

\section{ Discussions and conclusions}
   There are two more questions to be discussed. The first one is the status 
of the Witten index approach to this problem. Our statement is that the Witten index 
is {\it not} a suitable tool for studying the supermembrane spectrum.
   The reason is that, for systems with the continuous spectrum, the concept of 
index, i.e.  the number of unpaired zero-energy states, is poorly defined --- 
the gap between
the ground state and the excited states is absent here. Attempts to calculate 
Witten index in the systems not involving this gap may lead to
 rather puzzling results such as the fractional values for the index \cite{11}. 
Our early calculation of the Witten index in the $SU (2)$ SYM quantum mechanics 
performed in Ref. \cite{5} has an unclear status by the same reason.
   
Note in passing that the calculation of the index for supersymmetric nonchiral 
gauge theories  performed  in Ref. \cite{3} with the method based on the
periodic boundary conditions is also not well grounded. 
The reason for that is not the continuous spectrum 
(in the finite volume, the motion is finite), but the fact that there are regions 
in slow variables configuration space where the Born-Oppenheimer 
approximation is not applicable. This remark resolves the
apparent contradiction between Witten's result $I_W$
=  rank of
the group + 1 and the calculation of index based on the instanton calculus for 
higher orthogonal and exceptional groups 
(see the detailed discussion in Refs. \cite{6,12}).

    Finally, I would like to speculate on  possible ways 
to cure the continuous spectrum illness. We have seen earlier that the basic 
reason for this is the instability of supermembrane which tends to smear out 
emitting "needles" of zero area. Seemingly, this instability could be 
suppressed if the action would include terms preventing the membrane from 
bending. Such an action has been built in case of strings in ref. \cite{13},
 has 
been extended on the supersymmetric case in ref. \cite{14} and is known as 
"rigid string" action. It involves the square of extrinsic curvature of the 
world sheet and hence high curvature configurations of rigid string are 
suppressed.
   So one can suggest the program: to built up the analog of Polyakov action 
for supermembranes and look at the mass spectrum of its mass operator. 
Presumably, it is discrete.

\end{document}